\documentstyle[12pt,twoside,fleqn,espcrc1]{article}

\newcommand{\AmS}{{\protect\the\textfont2
  A\kern-.1667em\lower.5ex\hbox{M}\kern-.125emS}}

\title{Relativistic Mean Field calculations of nuclear properties in early 
stages of stellar collapse.}

\author{{F. K. Sutaria$^{\rm{a}}$, J. A. Sheikh} \address{Tata Institute of 
Fundamental Research (TIFR),\\ Bombay: 400 005, INDIA}%
and 
        A. Ray \address{Code 661, Goddard Space Flight Center, Greenbelt, 
MD20771, USA} 
 \thanks{NAS-NRC Senior Research Associate; on leave of absence from Tata 
Institute of Fundamental Research}}

\begin{document}
% typeset front matter
\maketitle

\begin{abstract}
 We use the Relativistic Mean Field (RMF) method to calculate 
properties of neutron rich, usually deformed nuclei, important for equation 
of state calculations and
which have significant abundance 
in the early stages of stellar collapse. We compare the results  
of our microscopic calculations  with existing cold nuclear equations of 
state based on macroscopic liquid drop model and the FRLDM model.
\end{abstract}

\section{Introduction}
    Spectroscopy of neutrinos produced by electron capture on neutron-rich 
f-p shell nuclei having significant abundance in the cores of nearby 
presupernova and collapsing stars, and emitted before neutrino trapping sets in 
(at core density of $\simeq 10^{12}$gm/cm$^3$), can yield useful
information on the physical conditions and nuclear composition of the core
because they stream freely through the overlying stellar matter 
without any further interactions$^{\cite{NIC94}}$.

The neutrino spectrum depends, among other factors, on nuclear abundances. 
The post-silicon burning ($\rho \geq 10^9 $ gm/cm and $T \geq 6 \times 10^{9}$ 
deg. K)
core composition consists of neutron rich f-p
shell nuclei whose abundance is controlled by nuclear statistical equilibrum.
The nuclear abundances thus depend on
the temperature dependant nuclear binding energies, and through the
neutron and proton fractions $X_n$ and $X_p$, on the neutron and proton
chemical potentials $\mu_n$ and $\mu_p$. 
The mass of the homologously
collapsing core and the strength of the hydrodynamic shock after stellar
core
bounce are determined by the 
lepton fraction and consequently by the electron capture rates$^{\cite{BBAL}}$. 
Since the e$^-$-capture
threshold too, depends on the nuclear binding energies it is neccessary to
calculate the chemical potentials and binding energies accurately, as functions 
of core temperature and density. 

Because the temperature and the drip neutron fraction
are relatively low, the cold   
nuclear equation of state (EOS) used should reproduce well
the laboratory values of nuclear binding energies and chemical potentials
for the neutron rich nuclei of interest. Since it must
also take into account the nuclear shell and pairing effects which 
persist upto temperatures $\simeq 0.5$ to 1 MeV, these quantum effects
have to be calculated using 
microscopic mean-field Hartree Fock or RMF methods. 
Thus, EOS based on the classical liquid-drop models$^{\cite{BBAL}-\cite{LS}}$, 
can only hold accurately beyond $\rho_{trapping} \simeq 10^{12}$ gm/cm$^3$
and T $\simeq 1$ MeV where nuclear shell and pairing effects are washed out.
For physical conditions before neutrino trapping sets in,
the cold EOS should go smoothly into the
high temperature macroscopic liquid drop model based EOS$^{\cite{LS}}$.
Since analytical forms of EOS are less computationally expensive than
detailed microscopic calculations when incorporated into hydrodynamic codes
to simulate stellar collapse and explosion,
we look for analytical models which would compare well with the microscopic
RMF calculations.
\section{ Existing Equations of State}

In general, all EOS developed so far approximate the ensemble of heavy nuclei 
and drip nucleons by a 
lattice consisting of a single heavy spherical 
nucleus (A, Z)
(which is the most bound system for a given core configuration) 
immersed in a sea of drip neutrons
of density $n_n$ and electrons of density $n_e$. 

 In the cold, incompressible liquid drop formalism, 
the nuclear matter energy of a single nucleus in a lattice, $W_N$,  
is given by$^{\cite{BBP},\cite{BBAL}}$ :
\begin{equation} 
  W_N( x, \rho_N, V_N, u) = W_{bulk} + 290x^2(1-x)^2 A^{2 \over3} 
 +\beta x^2 \rho_N^2
V_N^{5 \over 3} (1 -{3\over2} u^{1\over 3} + {1 \over 2} u)  
\label{Wn}
\end{equation} 
 where $x$ is the proton fraction , $\rho_N$ the nuclear density,  $V_N$ the 
nuclear
volume and $u$ the fraction of total volume occupied by nuclei ($u=\rho/\rho_N
$). This model accounts for the effects of drip neutrons and the nuclear
lattice, but not for the nuclear deformation or shell and pairing effects.
Minimising the energy of the system with respect to the nucleons bound in 
nuclei at fixed $n_NA$, $n_NZ$, $n_n$ and $n_N V_N$ gives the mass number 
of the (most bound) `mean' nucleus $A$. 
\section{Results}
We used the microscopic RMF approach with the Lagrangian set with non-linear 
self
interactions for the $\sigma $-meson to calculate nuclear properties for
isolated nuclei at zero temperatures, since it has been known to reproduce
the ground state properties of $\beta$-stable nuclei with sufficient accuracy.
%In this approach$^{\cite{JAS}}$, the nucleus is
%treated as an ensemble of Dirac Spinor nucleons interacting with mesonic
%and electromagnetic fields. 
The parameters that enter into the Lagrangian
include the nucleon mass $M_B$, the masses of the $\sigma$, $\omega$ and
$\rho$ mesons ($m_{\sigma}$, $m_{\omega}$, $m_{\rho}$) and the coupling
constants $g_\sigma$ $g_\omega$ and $g_\rho$. These are self consistently
determined by variational calculation.
The mesonic masses and coupling constants are treated as fixed parameters,
their values having been derived by fitting to ground state properties of
a few select spherical nuclei. The pairing is dealt within the BCS 
approximation and the initial values of the pairgaps were obtained from the 
odd-even mass differences, but it was found that to reproduce experimental
binding energies, it was neccessary to decrease pairgaps as the nuclear 
assymetry increased. 
Fig.(1) compares the experimental binding energies with the RMF values
for a range of nuclei in interest to this stage of collapse.  
Nuclear deformations are shown in Fig.(2) for spherical Ni and deformed Zn
isotopes.
The neutron chemical potential $\mu_n$ is obtained from the incompressible
liquid drop model as$^{\cite{Full82}}$: 
\begin{equation}
\mu_n = -16 + 125(0.5 -x) -125(0.5 -x)^2 -290 x^2 (1-x)^2 A^{-{1 \over 3}}
{(3 -7 x) \over 2(1-x)}
\label{incomp}
\end{equation}
%\newpag
%\mbox{}
%\flushbottom{This is a new text.}
\begin{figure}
%\vskip 16.5 true cm
\vspace*{16.5 cm}
\caption{Binding Energies vs. Neutron no. N for Mn, Fe, Co, Ni and Cu isotopes}
\vskip -0.8 true cm
\caption{Quadrupole($\beta_2$) deformation vs. Neutron no. N for Ni and Zn 
isotopes}
\vskip -0.8 true cm
\caption{Comparision of analytical models with RMF calculations for Ni isotopes}
\vskip -0.8 true cm
\caption{Comparision of analytical models with RMF calculations for Zn isotopes}
\vskip -0.8 true cm
\end{figure}
%\newpage
The first 3 terms in this expression come from the nuclear bulk energy, 
the last from coulomb and surface contributions to the nuclear energy.
 
Since the nuclear matter is being treated as an ensemble of isolated cold nuclei
with a low drip neutron fraction, we can also
derive a nuclear equation of state using the Finite Range Liquid Drop 
Model$^{\cite{MN}}$ (FRLDM). 
This gives the following expression for the volume 
contribution to  $\mu_n$ :
\begin{equation}
$$ \mu_n|_{vol} = -16.00 + 123.04(0.5 -x)^2 + 246.08 x (0.5 -x) \label{frldm}
\end{equation}
and an expression similar to that in eq.(\ref{incomp}) for the finite size
and coulomb effects.
Note that the bilinear term in eq.(\ref{frldm}) reduces to the linear term in 
eq.(\ref{incomp}) when nuclei in neighbourhood of
$x \simeq 0.5$ are considered.  
The results of the FRLDM model are compared with 
the RMF and 
values from eq.(\ref{incomp}) in Fig.(3) for
the spherical Ni systems and in Fig.(4) for the deformed Zn systems. We find 
that the FRLDM 
model reproduces better the RMF values than eq.(\ref{incomp}) even in the case
of spherical Ni nuclei. 
\subsection {Extension of the EOS to finite temperatures.}
At finite temperatures, the free energy $F$ of the system is modified 
to$^{\cite{Coop}}$
\begin{equation}
 F =  W_N - {a \over A}  {m^* \over m} T^2  \label{ftemp}
\end{equation} 
where $a$ is the nuclear level density parameter, and $m^*$ is the effective 
nucleon mass, which in general is a function of the nuclear temperature. 
We are using the RMF code to calculate the level density of the last 
filled orbitals to extract the coefficient of the temperature dependent
correction.

This work was carried
out under 8th Five Year Plan Project 8P-45 at T.I.F.R.


\begin{thebibliography}{99}

\bibitem{NIC94}F.K. Sutaria and A. Ray, Proceedings of Nuclei in Cosmos 
3$^{rd}$ International conference in Nuclear Astrophysics (1994) pg. 187, 
Ed. M. Busso and C. Raiteri.

\bibitem{BBP} G. Baym, H. A. Bethe and C. J. Pethick, 
Nucl.  Phys. {\bf A175} 225 (1971).
 
\bibitem{BBAL} H. A. Bethe, G. E. Brown, J. Applegate and J. M. Lattimer, 
Nucl. Phys. {\bf A324} 487 (1979).

\bibitem{BBCW} H. A. Bethe, G. E. Brown, J. Cooperstein and J. R. Wilson, 
Nucl. Phys. {\bf A403} 625 (1983).

\bibitem{Coop} J. Cooperstein, Nucl. Phys. {\bf A438} 722 (1985).

\bibitem{CB} J. Cooperstein and E. Baron, Supernova pg. 213-266 (1990), 
Ed. A. G. Petschek, Springer-Verlag. 

\bibitem{LS} J. M. Lattimer and F. D. Swesty, 
Nucl. Phys. {\bf A535}, 331-376, (1991). 

\bibitem{Full82}G. M. Fuller, 
Ap. J. {\bf 252} 741 (1982) 

\bibitem{MN} P. M\"oller, J. R. Nix, W. D. Myers and W. J. Swiatecki, 
At. Data and Nucl. Data tables, {\bf 59} 185-381 (1995). 

%\bibitem{Hix}W. R. Hix and F. -K. Thielemann,  Ap. J. {\bf 460}, 869-894,  
(1996).

\bibitem{JAS} J. A. Sheikh, J.P. Maharana and Y.K. Gambhir, 
Phys. ReV. C{\bf 48}, 192 (1993)

\end{thebibliography}
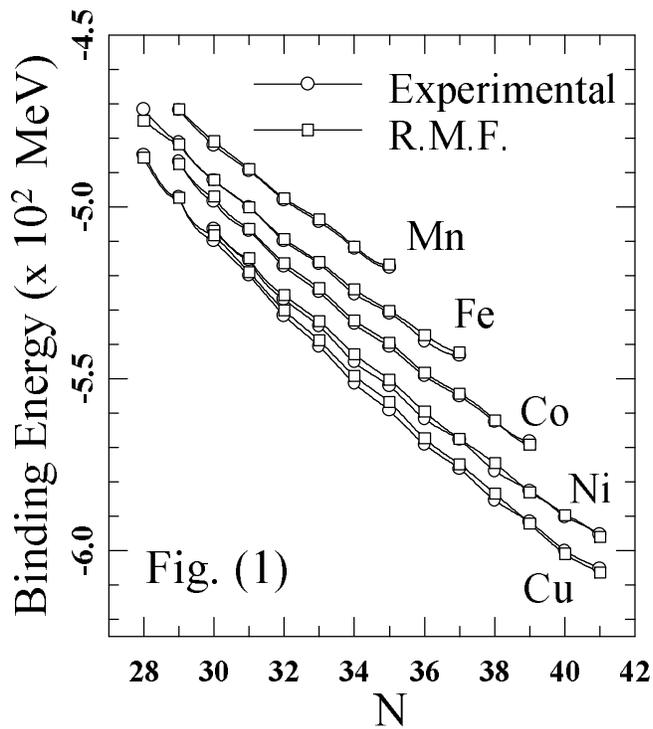
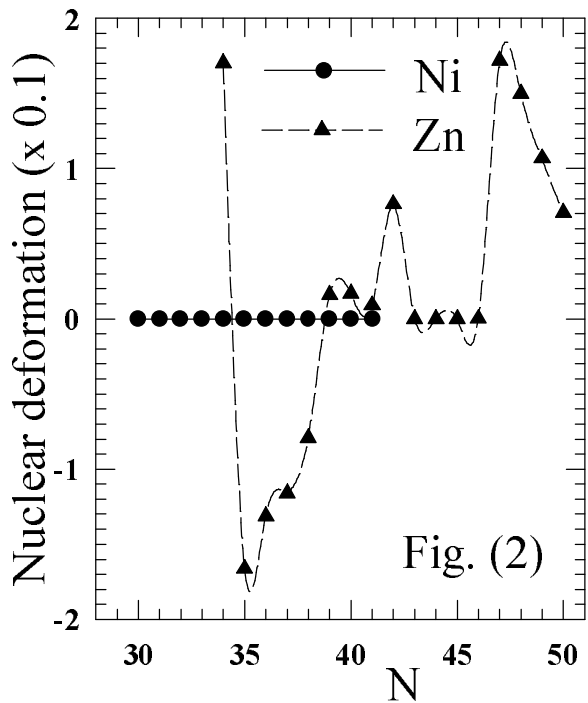
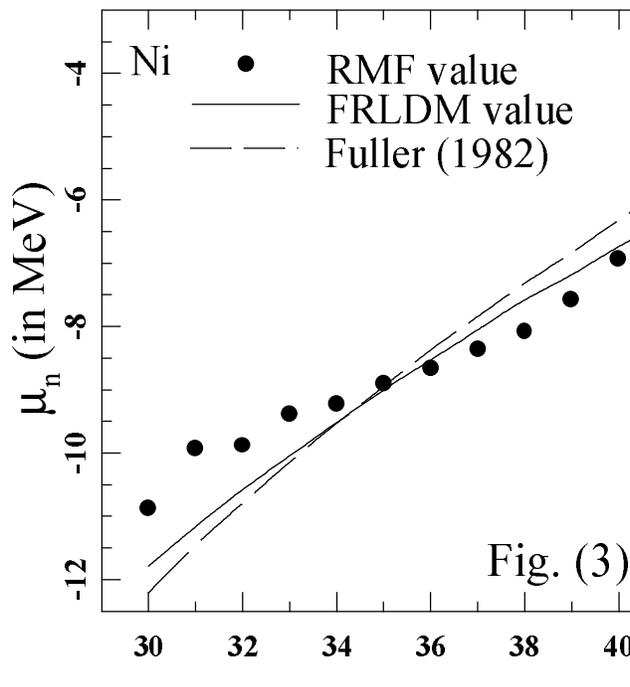
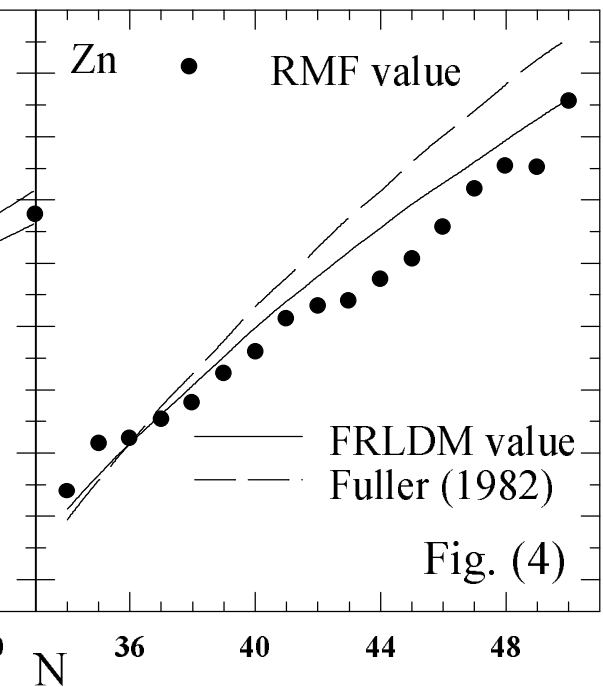
\end{document}